\documentclass[showpacs,10pt,twocolumn,prl]{revtex4-1}
\usepackage{amsmath}
\usepackage{amssymb}
\usepackage{graphics}
\usepackage{epsfig}
\usepackage{color}

\begin{document}
	
	
\title{Interstitial anionic electrons involved superconductivity and $T$-linear resistivity behavior in electride La$_{3}$In}
\author{Zhijun Tu$^{1,2,\dag}$, Peihan Sun$^{3,\dag}$, Pengcheng Ma$^{4,\dag}$, Hongrun Zhen$^{4}$, Shangjie Tian$^{5}$, Shouguo Wang$^{5}$, Tian Cui$^{4}$, Zhonghao Liu$^{4,*}$, Kai Liu$^{1,2,*}$, and Hechang Lei$^{1,2,*}$}
\affiliation{$^{1}$School of Physics and Beijing Key Laboratory of Optoelectronic Functional Materials \& MicroNano Devices, Renmin University of China, Beijing 100872, China\\
	$^{2}$Key Laboratory of Quantum State Construction and Manipulation (Ministry of Education), Renmin University of China, Beijing 100872, China\\
	$^{3}$Department of Physics, School of Science, Hebei University of Science and Technology, Shijiazhuang, Hebei  050018, China\\
	$^{4}$Institute of High-Pressure Physics and School of Physical Science and Technology, Ningbo University, Ningbo 315211, China\\
	$^{5}$Anhui Key Laboratory of Magnetic Functional Materials and Devices, School of Materials Science and Engineering, Anhui University, Hefei 230601, China
}
	
\date{\today}
	
\begin{abstract}
Electrides are unique materials because of the existence of interstitial anionic electrons (IAEs).
Due to these loosely bound IAEs and their strong interaction with the framework of cations, electrides can host superconductivity with rather high $T_{c}$, especially under high pressure, as predicted in theory. 
However, the experimental observations of superconductivity in electrides are very rare, let alone the detailed studies on intrinsic properties of single crystals. 
Here, we report the superconducting and normal-state properties of electride La$_{3}$In single crystals. 
La$_{3}$In shows a type-II superconductivity with $T_{c}\sim$ 9.4 K and a $T$-linear resistivity in a wide temperature range. 
Experimental measurements and theoretical calculations suggest that the relatively high $T_{c}$ could be ascribed to the high density of states around the Fermi level caused by short flat bands along R-M direction and the strong electron-phonon coupling, partially derived from the IAEs. 
Meanwhile, the $T$-linear resistivity may reflect the significant electronic correlation effect in this material. 
These findings will shed light on understanding the role of IAEs in superconductivity and open a promising way to explore high-temperature superconductors in electrides.
\end{abstract}
	

\maketitle
	
Electrides are a class of exotic compounds in which some electrons are localized at interstitial sites and behave as anions without nuclei.\cite{Dye1,Dye2}
Because interstitial anionic electrons (IAEs) are not bounded to speciﬁc atomic cores and they have significantly smaller mass when compared to normal anions, these loosely bound IAEs exhibit some unique properties, such as low work function, high hyperpolarizabilities, and very high conductivity.\cite{Toda,XuHL,Lee} 
These features enable electrides to be great useful in many applications. For example, the electrides can be a promising cathode material for organic light-emitting diodes and batteries as well as an efficient catalyst for ammonia synthesis.\cite{Kim1,HuJ,Kitano}
In addition, metallic electrides can also host superconductivity especially under high pressure.\cite{ZhaoZ,LiuZ1,LiuZ2,ZhaoY,GuoZ,WanZ,HuangHM,ZhangX,WangQ} Previous theoretical calculations have predicted that some of high-pressure electrides can possess the values of superconducting transition temperatures $T_{c}$'s beyond the McMillan limit or even close to the temperature of liquid nitrogen, such as $T_{c}\sim$ 49 K for $P$6/$mmm$ Li$_{5}$N at 150 GPa,\cite{WanZ} $\sim$ 54 K for $R\overline{3}$ Li$_{8}$Cs at 380 GPa,\cite{HuangHM} $\sim$ 73 K for $Fm$-3$m$ Li$_{8}$Au at 250 GPa,\cite{ZhangX} and $\sim$ 46 K for $Fm$-3$m$ Na$_{2}$H at 30 GPa.\cite{WangQ} 
These results arouse extensive research interest to explore possible high-temperature superconductors based on electrides. 
Although high-pressure electride superconductors have been predicted intensively, the experimental studies are scarce because of the harsh synthesis conditions under high pressure. As a prototype, the cI16-Li phase formed above 41 GPa (space group $I\overline{4}3d$) with  $T_{c}\sim$ 20 K is regraded as an electride superconductor.\cite{Shimizu,Struzhkin,Hosono1}
In comparison with the high-pressure superconducting electrides, some electrides  can exhibit superconductivity at ambient pressure. Actually, the first room-temperature and air-stable inorganic electride [Ca$_{24}$Al$_{28}$O$_{64}$]$^{4+}$$\cdot$4e$^{-}$ with zero-dimensional IAEs becomes a superconductor at $T_{c}\sim$ 0.4 K.\cite{Matsuishi,Miyakawa}
Additionally, Mn$_{5}$Si$_{3}$-type electride Nb$_{5}$Ir$_{3}$ with one-dimensional IAEs has been found to exhibit superconductivity with $T_{c}\sim$ 9.4 K,\cite{ZhangY} but whether the IAEs play a dominant role in the superconductivity of this material is still unknown.  

In addition to the superconducting properties, the electric resistivities of metallic materials at normal states also demonstrate diverse behaviours. 
For conventional metals, when the electron-phonon scattering process is dominant, the temperature dependence of electrical resistivity $\rho(T)$ usually shows a $T$-linear dependence at high-temperature region ($T^1$ relation) and crossovers to a $T^{n}$ behaviour with $n\sim$ 5 at low-temperature region.\cite{Ziman} On the other hand, if the electron-electron scattering is dominant, $\rho(T)$ will decrease quadratically with temperature ($T^2$ relation) at low-temperature region.\cite{Ziman}
However, strange metals can exhibit a $T$-linear-dependent $\rho(T)$ extended down to very low temperature even zero,\cite{Phillips} which is notoriously observed in high-temperature cuprate superconductors,\cite{Gurvitch} and also occur in heavy fermion materials, iron-based superconductors and magic-angle twisted bilayer graphene etc.\cite{Lohneysen,Doiron-Leyraud,Jaoui}
Thus, the $T$-linear $\rho(T)$ becomes a widespread phenomenon in strongly correlated metals. 
Some studies have attributed this behaviour to charge carriers Planckian-limit scattering at a rate given by $\hbar/\tau = \alpha k_{\rm B}T$, where $\alpha$ is a constant of order unity, $\hbar$ is the Planck constant and $k_{\rm B}$ is the Boltzmann constant.\cite{Hartnoll} Meanwhile, other viewpoints, such as antiferromagnetic spin ﬂuctuations\cite{Monthoux} and quantum criticality,\cite{Dagan} have also been proposed. 
However, the underlying origin of the $T$-linear resistivity is still under debate and has emerged as one of the major puzzles in the physics of metals.

In this work, we carried out a detailed study on La$_{3}$In single crystal and found that it is a type-II superconductor with $T_{c}\sim$ 9.4 K, while theoretical calculations suggest that there are IAEs located at the center of empty La$_{6}$ octahedra, i.e., La$_{3}$In can be regarded as an electride superconductor. 
Further analysis implies that the large density of states (DOS) at the Fermi level $E_{\rm F}$, partially originating from the IAEs, as well as the strong electron-phonon coupling lead to the enhanced $T_{c}$ when compared to the face centered cubic (FCC) La element ($T_{c}\sim$  5 - 6 K),\cite{Ziegler} which has high structural similarity to La$_{3}$In.
In addition, La$_{3}$In exists a $T$-linear $\rho(T)$ in a wide temperature range and linear energy dependence of scattering rates at normal state, suggesting certain electronic correlations in this system.	

\begin{figure}
\centerline{\includegraphics[scale=0.48]{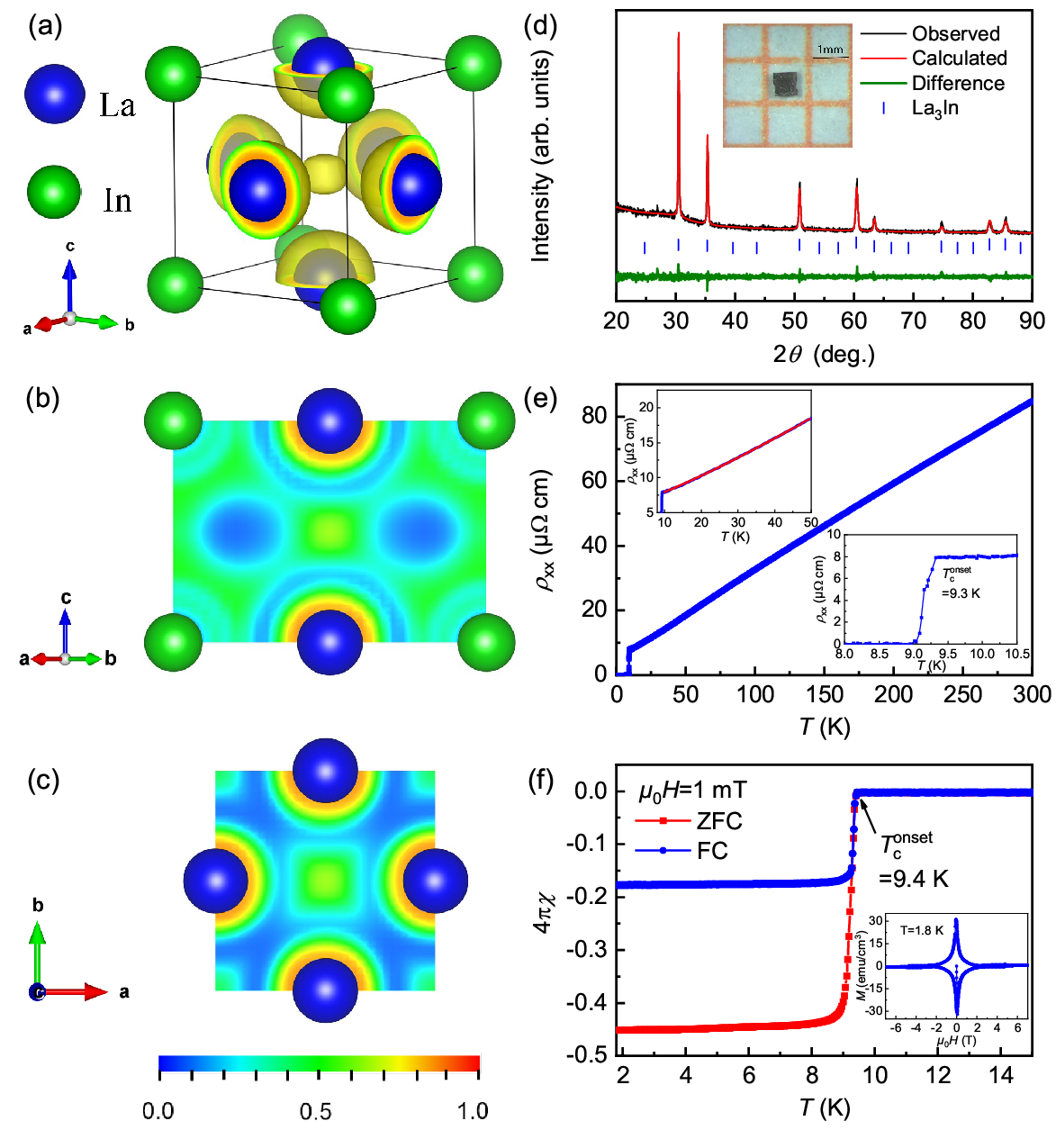}} \vspace*{-0.3cm}
\caption{The electron localization function (ELF) maps of La$_{3}$In with (a) the three-dimensional (3D) map, (b) and (c) the two-dimensional (2D) maps projected onto the (110) and (001) planes, respectively. The isosurface values set to 0.5. Blue and green balls represent La and In atoms, respectively. 
	(d) PXRD pattern of ground  La$_{3}$In single crystals. Inset: photo of typical La$_{3}$In single crystal on a 1 mm grid paper. 
	(e) Temperature dependence of $\rho_{xx}(T)$ at zero field for La$_3$In single crystal. Upper inset: the fit of normal state $\rho_{xx}(T)$ (red solid line) using the formula $\rho_{xx}=\rho_0+AT^n$. Lower Inset: the enlarged view of $\rho_{xx}(T)$ curve below 10.5 K. 
	(f) Temperature dependence of $4\pi\chi(T)$ measured at 1 mT with ZFC and FC modes. Inset: the magnetization loops for La$_3$In at $T=$ 1.8 K.}
\end{figure}

La$_3$In single crystals were grown by the La-flux method. The detailed methods of experimental characterizations and theoretical calculations are shown in Supplemental Material (SM).\cite{SM}
Figure 1(a) shows the crystal structure of La$_3$In, which is isostructural to Cu$_{3}$Au.\cite{ZhaoJT} From the perspective of unit cell, La atoms occupy the face-centered positions of the cubic structure, while In atoms occupy the vertex positions. This structure is closely related to the FCC-La in which all of In atoms are replace by La atoms.
Interestingly, six La atoms in La$_3$In form an empty octahedron at the body center of the cubic structure, which can host either interstitial atoms\cite{Gschneidner} or anionic electrons as shown below.
According to the stoichiometric ratio of atomic species in La$_3$In, we deduce the existence of excess electrons in this material. Figures 1(a)-1(c) show the respective three-dimensional (3D) map and two-dimensional (2D) maps projected onto the (110) and (001) planes for the electron localization functions (ELF).
As can be seen clearly, there are indeed partial electrons separated from the nuclei and confined at the interstitial site of La octahedron, forming the IAEs.\cite{Dye2,Park,Matsuishi,Hosono2} 
Hence, La$_3$In belongs to the well-known electride compounds.
Figure 1(d) shows the PXRD pattern of La$_3$In polycrystals. It can be well fitted using the $Pm$-3$m$ space group (No. 221) and the lattice parameter $a=$ 5.0785 (3) \AA, consistent with previous results.\cite{Garde}
Figure 1(e) shows the temperature dependence of electrical resistivity $\rho_{xx}(T)$ for La$_3$In from 300 K to 1.6 K at zero field.
The $\rho_{xx}(T)$ decreases with decreasing temperature, indicating the metallic behaviour of La$_3$In, and the residual resistivity ratio RRR (= $\rho$(300 K)/$\rho$(10 K)) is about 10.5.
Importantly, the $\rho_{xx}(T)$ curve exhibits a striking linear behaviour in a wide temperature range between 10 K and 300 K (upper inset of Fig. 1(e)).
It can be fitted very well by using the formula $\rho_{xx}(T)=\rho_0+AT^n$ with $n=$ 0.95(1), which is distinctively different from many of normal metals that usually show a $T^{2}$ or $T^{5}$ behaviour due to conventional electron-electron or electron-phonon scattering. 
Such linear temperature dependence of $\rho_{xx}(T)$ implies that there may be significant electron correlation effects in La$_{3}$In, leading to the obvious deviation from Fermi liquid behaviour.
On the other hand, with lowering temperature further, La$_{3}$In exhibits a superconducting transition and the onset superconducting transition temperature $T_{c}^{\rm{onset}}$ is about 9.3 K (lower inset of Fig. 1(e)), close to the reported values previously ($T_c\sim$ 9.7 K).\cite{Garde}
Correspondingly, there is a diamagnetic signal in magnetic susceptibility $4\pi\chi$ curves at 1 mT (Fig. 1(f)) and the $T_{c}^{\rm{onset}}$ is about 9.4 K, consistent with the value obtained from $\rho_{xx}(T)$ curve.
At $T=$ 1.8 K, the superconducting volume fraction estimated from the zero-field-cooling (ZFC)  curve is about 40 - 50 \%, confirming the bulk nature of superconductivity in La$_{3}$In.
Compared with the ZFC $4\pi\chi(T)$ curve, a relatively weak diamagnetic signal of FC $4\pi\chi(T)$ curve due to flux pinning effect suggests that La$_{3}$In should be a type-II superconductor.
The $M(\mu_0H)$ curve at 1.8 K shows obvious hysteresis (inset of Fig. 1(f)), further confirming the feature of  type-II superconductor for La$_3$In.

\begin{figure}
\centerline{\includegraphics[scale=0.46]{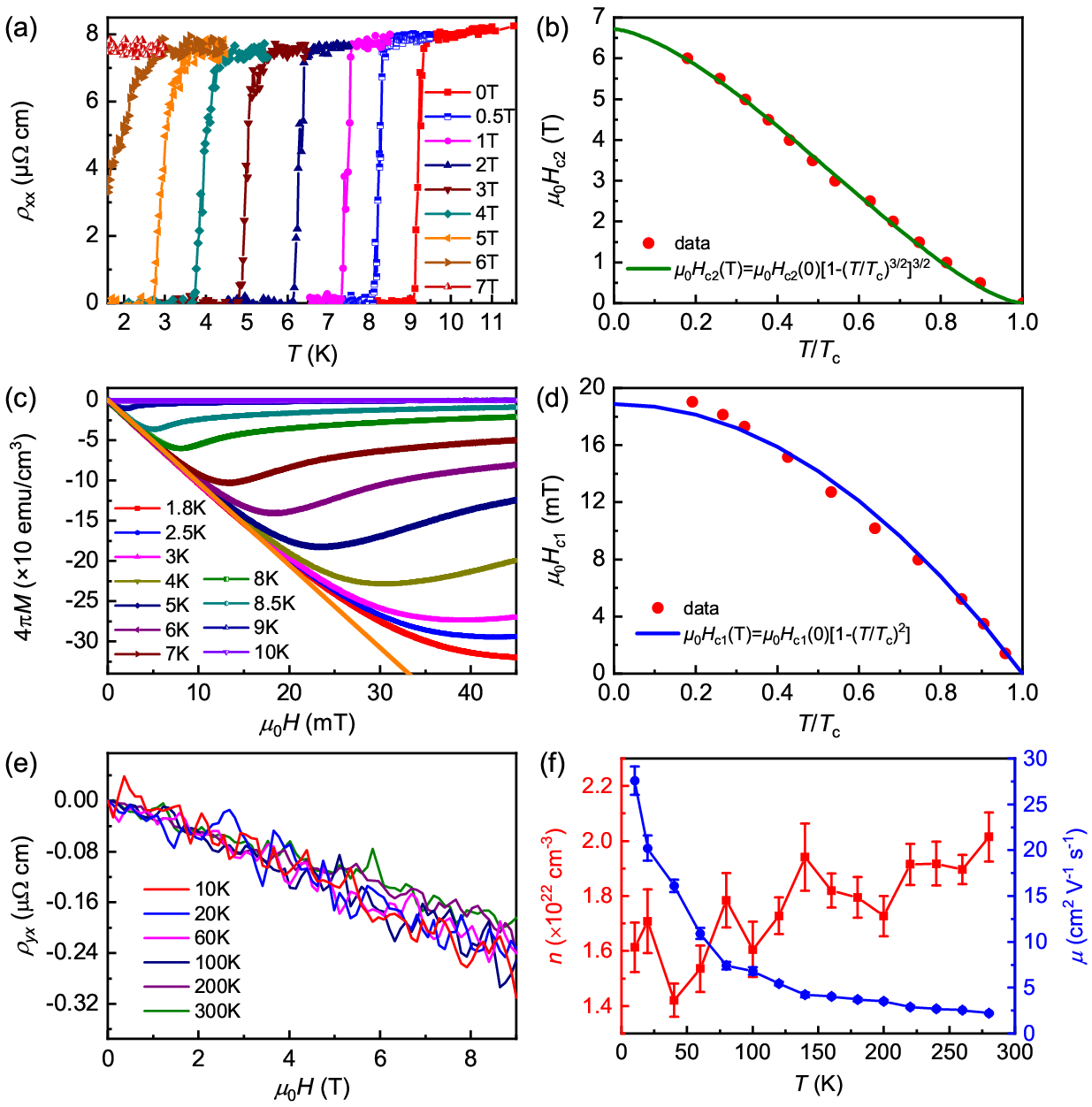}} \vspace*{-0.3cm}
\caption{(a) $\rho_{xx}(T)$ as a function of temperature in various applied magnetic fields up to 7 T. 
	(b) Temperature dependence of $\mu_0 H_{c2}(T)$. The green line represents the fit using the formula $\mu_0 H_{c2}(T)= \mu_0 H_{c2}(0)[1-(T/T_c)^{3/2}]^{3/2}$. 
	(c) Low-field parts of $4\pi M(\mu_0H_{\rm eff})$ curves at various temperatures below $T_{c}$. The orange line is the Meissner line. 
	(d) Temperature dependence of $\mu_0 H_{c1}(T)$. The blue line is the fit using the formula $\mu_0 H_{c1}(T)=\mu_0 H_{c1}(0)[1-(T/T_c)^2]$. (e) Field dependence of $\rho_{yx}(\mu _{0}H)$ at various temperatures. (f) Temperature dependence of derived $n(T)$ and $\mu(T)$.
}
\end{figure}

Figure 2(a) shows the $\rho_{xx}(T)$ curves at various magnetic fields up to 7 T. With increasing field, the $T_c$ shifts to lower temperature gradually but the width of superconducting transition only increases slightly, implying the relatively weak flux pinning effect in La$_{3}$In.
The upper critical field $\mu_0 H_{c2}$ is determined using the criterion of 50 \% normal state resistivity just above $T_c$ and summarized in Fig. 2(b).
It can be seen that the  $\mu_0 H_{c2}(T)$ increases with decreasing temperature with a slope $\frac{dH_{c2}}{dT}|_{T=T_{c0}}$= -1.05 T K$^{-1}$, where $T_{c,0}$ is the superconducting transition temperature at zero field.
The $\mu_0 H_{c2}(T)$ can be fitted using the empirical formula $\mu_0 H_{c2}(T)=\mu_0 H_{c2}(0)[1-(T/T_c)^{3/2}]^{3/2}$ and the fitted value of zero-field upper critical field $\mu_0 H_{c2}(0)$ is 6.72(4) T.
This is much smaller than the weak-coupling Pauli paramagnetically limited field $\mu_0$$H_{c2}^P(0)$ = 1.84$T_c$ = 17.11 T, implying that Pauli paramagnetic effect is negligible and the orbital depairing effect should be dominant in La$_3$In.
Using the fitted $\mu_0 H_{c2}(0)$, the calculated Ginzburg-Landau coherence length $\xi_{\rm GL}(0)$ from the equation $\xi_{\rm GL} = \sqrt{\Phi_{0}/2\pi\mu_{0}H_{c2}}$ ($\Phi_0=h/2e$ is quantum flux) is 70.0(3) \AA.

In order to obtain the lower critical field $\mu_0 H_{c1}(T)$, the $4\pi M$ as a function of field at various temperatures below $T_c$ were measured (Fig. 2(c)).
The $4\pi M(\mu_0H_{\rm eff})$ curves exhibit linear behaviour at low-field region and the slopes are very close to -1, corresponding to the Meissner line (orange line) when the effective field $\mu_0 H_{\rm eff}$ is calculated using the formula $\mu_0 H_{\rm eff}= \mu_0 H_{\rm ext}-N_dM$, where $\mu_0$$H_{\rm ext}$ is the external field and $N_d$ is the demagnetization factor.
The values of $\mu_0 H_{c1}(T)$ are determined from the fields where the $4\pi M(\mu_0H_{\rm eff})$ curves deviate from the linearly field-dependent behaviour, i.e., the magnetic flux starts to penetrate into the sample.
As shown in Fig. 2(d), the $\mu_0$$H_{c1}(T)$ increases with decreasing temperature and it can be fitted well using the equation $\mu_0 H_{c1}(T)= \mu_0 H_{c1}(0)[1-(T/T_c)^2]$, where $\mu_0 H_{c1}(0)$ is the lower critical field at $T=$ 0 K. The fitted $\mu_0 H_{c1}(0)$ is 18.9(3) mT.
The zero-temperature penetration depth $\lambda_{\rm GL}(0)$ can be evaluated using the formula $\mu_0 H_{c1}=\frac{\Phi_0}{4\pi\lambda_{\rm GL}^2}{\rm ln}\frac{\lambda_{\rm GL}}{\xi_{\rm GL}}$, and it is 1660(26) \AA.
Correspondingly, the calculated Ginzburg-Landau parameter $\kappa_{\rm GL} (= \lambda_{\rm GL}/\xi_{\rm GL}$) is 23.7(3), further confirming that La$_3$In is a type-II superconductor.
The zero-temperature thermodynamic critical field $\mu_0 H_{c}(0)$ can be calculated using the formula $\mu_0 H_{c}=\mu_0 H_{c2}/(\sqrt{2}\kappa_{\rm GL})$ and it yields $\mu_0 H_{c}(0)=$ 201(2) mT for La$_3$In.
The electron-phonon coupling constant $\lambda_{\rm e-ph}$ can be obtained from the McMillan equation $\lambda_{\rm e-ph}=\frac{\mu^{\ast}{\rm ln}(1.45T_{c}/\Theta _{D})-1.04}{1.04+{\rm ln}(1.45T_{c}/\Theta_{D})(1-0.62\mu^{\ast})}$. When assuming the Coulomb pseudopotential $\mu^{\ast}$ is about 0.15, using above values of $T_{c}^{\rm onset}$ and the reported $\Theta_{D}=$ 170 K,\cite{Ravindran} the value of $\lambda _{\rm e-ph}$ is determined to be 1.13(1), implying the strongly coupled BCS superconductivity in La$_{3}$In.
Figure 2(e) shows the magnetic field dependence of Hall resistivity $\rho_{yx}(\mu _{0}H)$ at different temperatures. All of curves  exhibit linear behaviours with negative slopes, suggesting that electrons dominate the transport in La$_3$In.
Combined with $\rho_{xx}(T)$ data, the carrier concentration $n(T)$ (red square) and mobility $\mu(T)$ (blue circle) as functions of temperature can be obtained using one band model (Fig. 2(f)).
The $n(T)$ is about 1.4 - 2.0 $\times$10$^{22}$ cm$^{-3}$ with a weak temperature dependence when the $\mu(T)$ decreases gradually with increasing temperature from 27.6(15) cm$^2$ V$^{-1}$ s$^{-1}$ at 10 K to 2.2(1) cm$^2$ V$^{-1}$ s$^{-1}$ at 280 K. 

\begin{figure}
\centerline{\includegraphics[scale=0.60]{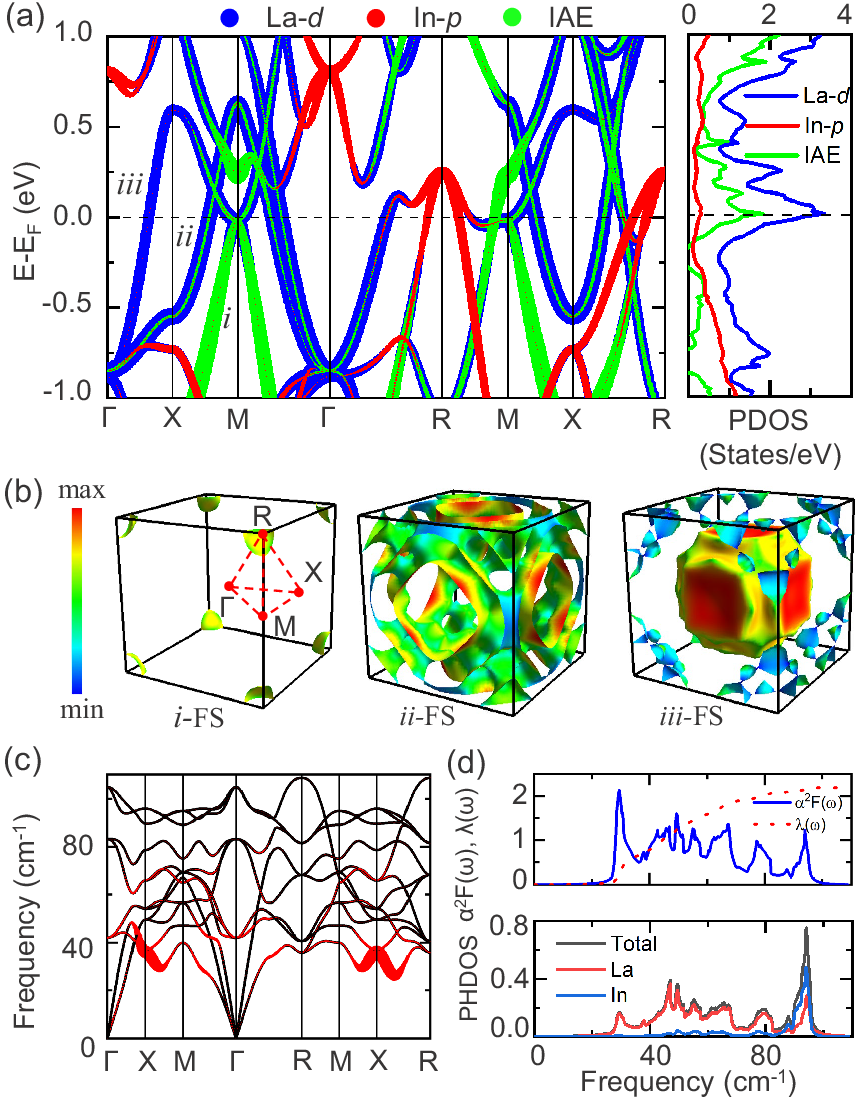}} \vspace*{-0.3cm}
\caption{(a) The orbital-resolved electronic band structure and partial density of states (PDOS). 
	(b) The Fermi-velocity-projected Fermi surface sheets corresponding to the bands \textit{i}, \textit{ii}, and \textit{iii} in (a), respectively. 
	(c) Phonon dispersion curves. The size of red dots represents the electron-phonon coupling (EPC) strength $\lambda_{{\bf q}\nu}$. 
	(d) The Eliashberg spectral function $\alpha^2F(\omega)$, frequency-dependent EPC constant $\lambda(\omega)$ and phonon density of states (PHDOS).
}
\end{figure}

Figure 3(a) shows the electronic band structure and the partial density of states (PDOS) of La$_{3}$In calculated without the spin-orbital coupling (SOC). 
There are three bands, labeled \textit{i}, \textit{ii}, and \textit{iii}, crossing the Fermi level $E_{\rm F}$, among which there are short flat bands along the R-M path of Brillouin zone (BZ). Based on the orbital weight analysis, these short flat bands are dominated by the La 5$d$ electrons and the IAEs. 
The Fermi surface (FS) sheets of bands \textit{i}, \textit{ii}, and \textit{iii} are correspondingly shown in Fig. 4(b), on which the Fermi velocities are displayed with the color scales, where the red and blue colors represent the highest and zero Fermi velocities, respectively. 
It can be seen that the electron-type FS of band \textit{iii} centered at $\Gamma$ point of BZ is much larger than other FSs and it also has a relatively high Fermi velocity, thus this electron band will contribute to the transport properties, which is in agreement with the results of Hall measurement (Fig. 2(e)).
It is clear that the FS of band \textit{ii} has a much lower Fermi velocity around the $M$ point, which is favorable for strong electron-phonon coupling (EPC).\cite{lower-velocity1,lower-velocity2}
According to the PDOS shown in Fig. 3(a), it is found that the electronic states near $E_{\rm F}$ are mainly contributed by the La 5$d$ orbitals and the IAEs, which are consistent with the results of band structure. 
More importantly, there is a van Hove singularity around $E_{\rm F}$, which is associated with the partial ﬂat bands around M point and results in a large DOS of 2.15 states/eV per atom at $E_{\rm F}$. Compared with the value of 1.63 states/eV per atom in FCC-La with $T_{c}\sim$  5 - 6 K,\cite{Ziegler,Bagci} it suggests that the large DOS at $E_{\rm F}$ ($N(E_{\rm F})$) for La$_{3}$In could lead to an enhancement of superconductivity.

To better understand the superconducting properties of La$_{3}$In, we subsequently performed the EPC calculations. 
The calculated total EPC constant $\lambda$ is 2.23, indicating a strong EPC in La$_{3}$In compared with that of FCC-La ($\lambda=$ 1.06).\cite{Bagci} 
This larger theoretical value of $\lambda$ than that derived from experimental results (1.13) could be ascribed to the difference between $\omega_{\rm log}$ ($\sim$ 67 K) and Debye temperature $\Theta_{D}$ ($\sim$ 170 K).\cite{Ravindran}
When $\lambda$ is larger than 1.3, one needs consider the correction factors $f_1$ and $f_2$, which are calculated to be 1.15 and 1.04, to obtain the superconducting temperature.\cite{Allen1,Allen2} These lead to a superconducting $T_c$ of 10.66 K based on the McMillan-Allen-Dynes formula,\cite{Allen1} which agrees well with the measured result (Fig. 1(d)). 
From the momentum- and mode-resolved EPC parameter $\lambda_{{\bf q}\nu}$ (Fig. 3(c)), it can be seen that the largest contribution comes from the acoustic branches around X point, which results in a high peak (around 30 cm$^{-1}$) in the Eliashberg spectral function $\alpha^2F(\omega)$ (the top part of Fig. 3(d)). 
Such phonon mode is closely related to $z$-axial vibration of La atoms in the La-In layer.
Combined the phonon density of states (PHDOS) and the frequency-dependent EPC parameters $\lambda(\omega)$ (Fig. 3(d)), we conclude that the La vibrations play a dominant role in the superconductivity. 
In short, we propose that the enhanced superconductivity in La$_{3}$In, compared with that of FCC-La,\cite{Bagci} is mainly due to the high electronic states at the Fermi level and the strong EPC strength induced by the low-frequency phonon modes.

\begin{figure*}
\centerline{\includegraphics[scale=0.65]{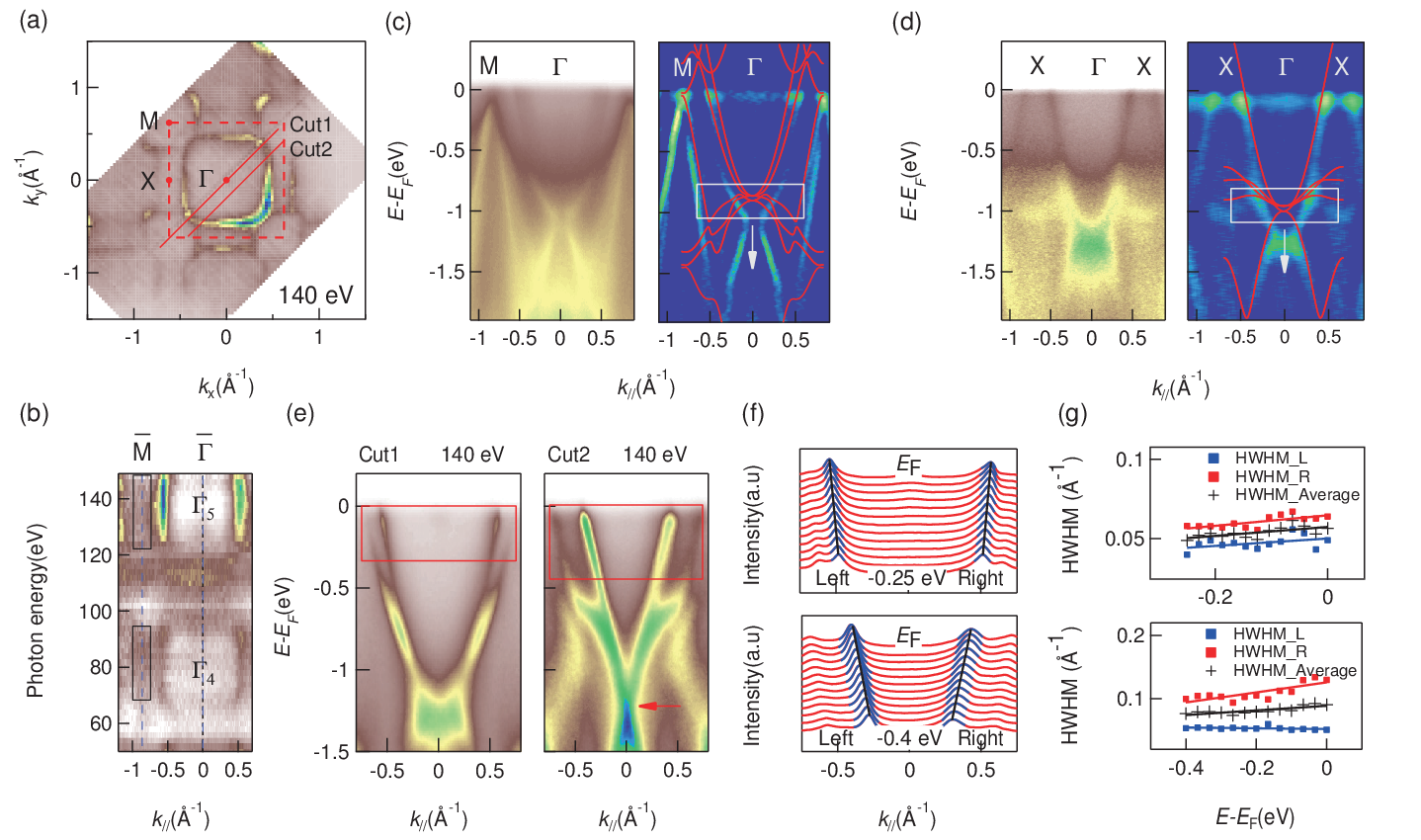}} \vspace*{-0.3cm}
\caption{(a) Integrated intensity plots at $E_{\rm F}$ $\pm$ 10 meV taken by the photon energy of 140 eV. The red frames indicate the 2D BZ projected on the (001) surface. 
	(b) Integrated intensity plots on the \emph{k$_{\parallel}$}-\emph{k$_z$} plane taken at $E_{\rm F}$. Photon energies used for the measurements were from 50 to 148 eV. 
	(c) and (d) Intensity plots and corresponding second derivative plots along the $\Gamma$-$M$ and $\Gamma$-$X$ directions, respectively. The red solid lines on the corresponding second derivative plots are DFT-calculated bands without any renormalizations and shifts. 
	(e) Intensity plots taken by 140-eV photos along the cut1 and cut2 in (a). 
	(f) The momentum distribution curves (MDCs) of (e). The linear bands are fitted by using the Lorentz function, as indicated by black solid curves. 
	(g) Energy-dependent scattering rate obtained from the half width at half maximum (HWHM) of the fitted MDCs in (f). Red squares are for the left branch, blue ones are for the right branch, and black crosses are for the average value of both branches. The data are linearly fitted as indicated by the lines with different colors.
}
\end{figure*}

Next, we performed angle-resolved photoemission spectroscopy (ARPES) measurements to identify band structures near $E_{\rm F}$ and confirmed the anomalous scattering rate correlated with the non-Fermi liquid behaviour. 
In Fig. 4(a), the FSs taken by a photon energy of 140 eV show big electron-like FSs at the BZ center as predicted by the calculation. 
However, due to the matrix elements effects, the calculated bands at the $M$ points were not observed by the 140-eV photons. We therefore have carried out measurements using various photon energies ranging from 50 to 148 eV to investigate $k_{z}$ dispersion along the $\Gamma$-$M$ lines in detail. 
Fig. 4(b) reveals the intensity as a function of phonon energy along $\Gamma$-$M$ taken at $E_{\rm F}$. According to the periodicity of intensity and free-electron final state model,\cite{kz} we found that $hv=$ 80 and 140 eV are close to the $\Gamma$ points. 
When using around 140-eV photons, the intensity of a electron-like pocket surrounding $\Gamma$ is enhanced and the intensity at the $M$ points is suppressed. In contrast, using around 80-eV photons, the intensity at $\Gamma$ and \emph{M} is almost the same. 
The band structures along the $\Gamma$-$M$ and $\Gamma$-$X$ directions are shown in Figs. 4(c) and 4(d). 
The calculated bands on the corresponding second derivative plots well match the mainly ARPES dispersive features without any renormalizations, except for the bands near -1 eV below $E_{\rm F}$ needing to be shifted down as marked in the figures which could be explained in terms of interband coupling and correlation effects.\cite{kz}

Interestingly, the electron-type FS surrounding the BZ center exhibits linear dispersions, which can be more clearly seen by using 140-eV photons as shown in Fig. 4(e) and the corresponding momentum distribution curves (MDCs) (Fig. 4(f)).
The scattering rates obtained from the fitted MDCs vary linearly with the binding energies up to 0.25 and 0.4 eV along the Cut1 and Cut2 lines, respectively. Despite the equal theoretical values of the half-width at half maximum (HWHM) on the right and left branches of the linear bands, some points deviate from the lines with opposite tendencies due to the effect of the intensity of near bands. The average values are even better for the linear relationship, especially for the MDCs of Cut2. A cone-like feature is formed at about -1.2 eV below $E\rm_F$ as marked at Cut2 in the Fig. 4(e). Anomalous scattering rates are inconsistent with the 3D Fermi-liquid theory ($1/\tau \propto \omega^2$) or the 2D Fermi-liquid theory [$1/\tau \propto (\omega^2/\epsilon_F)\ln(4\epsilon_F/\omega)$].\cite{Non_FL} Similar linear scattering rates are observed in iron pnictides,\cite{SM_JP,FB_JP} consistent with non-Fermi liquid behaviours observed in the resistivity. The non-Fermi liquid behaviours in these materials are possibly associated with correlation effects induced by high density states saddle points or partial flat bands at  $E\rm_F$, as also demonstrated in the calculated PDOS and band structure (Fig. 3(a)). 

When compared to FCC-La, the higher DOS around $E\rm_F$ in La$_{3}$In mainly originates from the La 5$d$ and IAE band near the $M$ point of BZ. 
Moreover, the proportion of IAE near $E\rm_F$ in the latter increases significantly. Therefore, we speculate that IAE is also involved in the phenomenon of enhanced superconductivity.
Such high DOS could also lead to certain electronic correlations and corresponding $T$ linear resistivity behaviour in this material.
In addition, there may be a strong EPC between IAEs at the center of La$_{6}$ octahedra and the $z$-axial vibration of apical La atoms of that La$_{6}$ octahedra. 
Unlike traditional anions characterized by strong electron localization, the IAEs usually exhibit weaker localization and higher activity.\cite{Lee} This feature can facilitate their hybridization with atomic valence orbitals and strengthening their coupling with the vibrations of surrounding La atoms.\cite{GuoZ,Miao,Racioppi} 
Similar enhancement of EPC have also been proposed in the Li$_{8}$Au where the IAEs are located at the cubic Li$_{8}$ cage centers and the IAEs induce strong coupling to the vibration of Li$_{8}$ cages enhancing the phonon mediated superconductivity.\cite{ZhangX}

For the $T$-linear behaviour of resistivity, usually, this phenomenon appears when $T\geq \Theta_{D}/5$ when the electron-phonon scattering dominates. As mentioned above, the $\Theta_{D}$ of La$_{3}$In is about 170 K, meaning that the $T$-linear behaviour should appear above 34 K. It is much higher than 10 K observed in La$_{3}$In.
Moreover, it is noted that for materials with small Fermi momentum $k_{\rm F}$, i.e., low carrier density $n$ since $k_{\rm F}\sim n^{1/d}$ with the system dimensionality of $d$, $\Theta_{D}$ should be replaced by $T_{\rm BG} = 2\hbar k_{\rm F}v$, where $v$ is the phonon velocity.\cite{Hwang}
Due to $T_{\rm BG}\sim n^{1/d}$, $T_{\rm BG}$ could be much less than $\Theta_{D}$ in low-density metals in principle, leading to the $T$-linear behaviour persisting to low temperature. 
But for La$_{3}$In, the $n$ is about  10$^{22}$ cm$^{-3}$ with a rather big FS around $\Gamma$ point of BZ, it seems unlikely to satisfy the condition of $T_{\rm BG}\leq\Theta_{D}$.
Thus, electron-phonon scattering might not be enough to  explain $T$-linear resistivity in the low-temperature region. Moreover, combined with the linear energy dependence of scattering rates observed from ARPES measurements, it suggests that such phenomenon at low temperature may have certain similarity to strongly correlated electronic systems like high temperature cuprate or iron-based superconductors. Nevertheless, unlike unconventional superconductors, the spin ﬂuctuations are absent in La$_{3}$In, which provides a constrain on the uniﬁed theoretical model in describing the $T$-linear resistivity. 

In summary, our present studies propose that La$_{3}$In should be an superconducting electride with rather high $T_{c}$, which are closely related to the high DOS, strong EPC, and partial ﬂat band near  $E\rm_F$ that involve both La 5$d$ electrons and IAEs.
In addition, the $T$-linear resistivity persists down to 10 K implies that besides electron-phonon interaction, the electronic correlation effects may also play an important role in the physical properties of La$_{3}$In.
In light of this work, we expect superconducting electride materials with IAEs can offer a new platform to explore exotic and/or high temperature superconductivity as well as to study the unique EPC and other electronic correlation phenomena.

This work was supported by the National Key R\&D Program of China (Grants No. 2023YFA1406500, 2022YFA1403800, 2022YFA1403103 and 2022YFB3608000), the National Natural Science Foundation of China (Grants No. 12274459, 12222413, and 12174443), the Beijing Natural Science Foundation (Grant No. Z200005), the Natural Science Foundation of Shanghai (Grants No. 23ZR1482200 and No. 22ZR1473300), and the funding of Ningbo Yongjiang Talent Program and Ningbo University. Computational resources were provided by the Physical Laboratory of High-Performance Computing at Renmin University of China.

$^{\dag}$ Z.J.T, P.H.S, and P.C.M contributed equally to this work.

$\ast$ Corresponding authors: liuzhonghao@nbu.edu.cn (Z. H. Liu); kliu@ruc.edu.cn (K. Liu); hlei@ruc.edu.cn (H. C. Lei).

\end{document}